\documentclass[twocolumn,floats,superscriptaddress,prl]{revtex4}

\usepackage[pdftex]{graphicx}
\usepackage{amsmath}
\usepackage{amsfonts}
\usepackage{mathrsfs}
\usepackage[utf8]{inputenc}
\usepackage{color}
\usepackage{xcolor}

\usepackage[normalem]{ulem}

\renewcommand\vec{\boldsymbol}

\begin{document}

\title{Defect Line Coarsening and Refinement in Active Nematics}

 \author{Nika Kralj}
 
 \affiliation{Faculty of Mathematics and Physics, University of
   Ljubljana, Jadranska 19, 1000 Ljubljana, Slovenia}

 \author{Miha Ravnik}
 
 \affiliation{Faculty of Mathematics and Physics, University of
   Ljubljana, Jadranska 19, 1000 Ljubljana, Slovenia}
 
 \affiliation{Condensed Matter Physics Department, J. Stefan Institute, Jamova 39, 1000 Ljubljana,
   Slovenia}
   
 \author{Žiga Kos}
 \email{ziga.kos@fmf.uni-lj.si}
 \affiliation{Faculty of Mathematics and Physics, University of
   Ljubljana, Jadranska 19, 1000 Ljubljana, Slovenia}
   \affiliation{Condensed Matter Physics Department, J. Stefan Institute, Jamova 39, 1000 Ljubljana,
   Slovenia}
   \affiliation{International Institute for Sustainability with Knotted Chiral Meta Matter, Hiroshima University, Higashihiroshima 739-8511, Japan}

\date{\today}

\begin{abstract}
    Active matter is naturally out of equilibrium which results in the emergence of diverse dynamic steady states, including the omnipresent chaotic state known as the active turbulence. However, much less is known how active systems dynamically depart out of these configurations, such as get excited or damped to a different dynamic steady state. In this Letter, we demonstrate the coarsening and refinement dynamics of topological defect lines in three-dimensional active nematic turbulence. Specifically, using theory and numerical modelling, we are able to predict the evolution of the active defect density away from the steady state due to time-dependent activity or viscoelastic material properties, establishing a single length scale phenomenological description of defect line coarsening/refinement in a three-dimensional active nematic. The approach is first applied to growth dynamics of a single active defect loop, and then to a full three-dimensional active defect network. More generally, this work provides insight into the general coarsening phenomena between dynamical regimes in 3D active matter, with a possible analogy in other physical systems.
\end{abstract}

\maketitle

Active matter systems are distinctly non-equilibrium in nature, but regularly form diverse dynamic steady states~\cite{GompperG_JPhysCondensMatter32_2020, MarchettiMC_RevModPhys85_2013}. Much like passive systems that evolve over time to reach equilibrium upon a stimulus, active systems can evolve into new or different dynamic steady states and  coarsening is observed as the major transitional mechanism~\cite{GeyerD_PhysRevX9_2019,RednerGS_PhysRevLett110_2013,FaddaF_arXiv220305213condmatsoft_2022,ChakrabortyS_JChemPhys153_2020,GonnellaG_ComptesRendusPhysique16_2015}.
Density correlation functions during coarsening separate active suspensions into classes~\cite{DeyS_PhysRevLett108_2012} with structure functions commonly deviating from the Porod's law that is generally expected for systems relaxing towards equilibrium~\cite{DeyS_PhysRevLett108_2012}.
In active binary fluids, the initial length scale dynamics during coarsening is reported to follow the same time dependence as for passive fluids~\cite{BertiS_PhysRevLett95_2005,SabrinaS_SoftMatter11_2015}, while at larger scales activity takes over and eventually a dynamic steady state is established~\cite{BertiS_PhysRevLett95_2005}.
Coarsening was observed also 
in two-dimensional dry active nematics and is based on annihilation of half-integer defect pairs~\cite{MishraS_PhilTransRSocA372_2014}.

Active nematics are a class of active materials, which exhibit apolar orientational order along the director $\vec{n}$, with  material examples including microtubule mixtures and bacterial suspensions
~\cite{MarchettiMC_RevModPhys85_2013, SanchezT_Nature491_2012, WensinkHH_ProcNatlAcadSci109_2012,AlertR_AnnuRevCondensMatterPhys13_2022,HardouinJ_SoftMatter16_2020}. In three dimensions, bulk active nematics form the dynamic steady state called active turbulence, which at the structural level is a dynamic rewiring network of defect lines and loops~\cite{DuclosG_Science367_2020,UrzayJ_JFluidMech822_2017,KrajnikZ_SoftMatter16_2020}, driven by the anisotropic active stress~\cite{HatwalneY_PhysRevLett92_2004,VoituriezR_EurophysLett70_2005}.
Recently, it was shown that advective terms are suppressed in active fluids~\cite{CarenzaLN_EPL132_2020,AlertR_NaturePhys16_2020} and in the steady state the energy injection is exactly matched by the viscous dissipation at each scale~\cite{AlertR_NaturePhys16_2020}.
Defect line segments are driven by the self-propulsion velocity depending on their local director profile, leading the defect loops to grow, shrink, and buckle in time~\cite{BinyshJ_PhysRevLett124_2020,LongC_SoftMatter17_2021}.

Beyond the active matter, the phase ordering kinetics through coarsening exhibits universal behaviour across a range of physical systems, as underlain with the fundamental role of the topological defects within the order parameter field~\cite{BrayAJ_AdvPhys51_2010}. Universal rules for phase kinetics are typically obtained through energetic arguments~\cite{BrayAJ_AdvPhys51_2010}, which opens a question what novel insights active matter energetics as an emergent field can provide. 
Notably, the universal of defects in the shape of lines and loops --- first described by Kibble for cosmic strings~\cite{KibbleTWB_JPhysAMathGen9_1976} and later predicted by Zurek for superfluid helium~\cite{ZurekW_PhysRep276_1996} --- was  experimentally observed in (passive, i.e. not active) nematic liquid crystals~\cite{ChuangI_Science251_1991}.

In this Letter, we show transitional dynamics from initial configurations towards a dynamic steady state and also between dynamic steady states of 3D active nematic turbulence, as distinctly determined by the coarsening and refinement of a  network of topological defect lines and loops. We  construct an analytical model of the collapse or growth of a single defect loop and then generalise it to the coarsening/refinement of the full 3D defect network. The approach  provides analytic insight into the effective phase ordering kinetics towards dynamic steady states, as triggered by changes in the main material parameters, such as activity or even nematic elasticity and viscosity. While the notion of self-propelled defects is unique for active nematics, the demonstrated  coarsening-refinement indicates possible universal behaviour applicable to different physical systems, including cosmic string dynamics.

Active nematics are described by the experimentally supported~\cite{DuclosG_Science367_2020,WoodhouseFG_PhysRevLett109_2012} mesoscopic active nematodynamic formulation~\cite{CoparS_PhysRevX9_2019,CarenzaLN_ProcNatlAcadSci116_2019}.
The approach is based on the coupled dynamics of the two main fields --- the velocity field $\vec{v}$ and the nematic order parameter tensor Q with the director $\vec{n}$  as the main eigenvector. Flow field is determined by the active propulsion due to the active stress that is proportional to $\text{Q}$~\cite{HatwalneY_PhysRevLett92_2004}, and by the viscous coupling to the nematic order, whereas the dynamics of  Q is determined  by the interplay between the dissipative relaxation towards the equilibrium and coupling to the material flow (SI).  
We solve this model by using a hybrid lattice Boltzmann algorithm with the results given in units of mesh resolution $\Delta x$, tensorial elastic constant $L$, and rotational viscosity $\Gamma$. 
Such numerical approach was shown to reproduce different structural and dynamical features of multiple experimental two-dimensional~\cite{DoostmohammadiA_NatCommun9_2018,ZhangR_NatCommun7_2016} and three-dimensional active nematic systems~\cite{DuclosG_Science367_2020,KrajnikZ_SoftMatter16_2020,BinyshJ_PhysRevLett124_2020}.

\begin{figure}[t!]
  \includegraphics[width=\columnwidth]{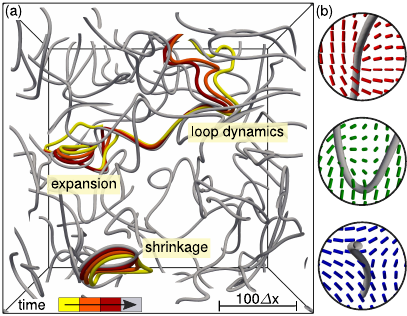}
  \caption{ Coarsening dynamics of active nematic turbulence. (a) A snapshot of a defect line network (gray) during coarsening process. Additionally, selected defect line segments are drawn in different colours at earlier time intervals  $50\,\Delta x^2/(\Gamma L)$ apart. (b) The director field rotates for an angle of $\pi$ around the defect lines: from $+1/2$ profiles (red), to twist (green) and $-1/2$ profiles (blue). 
  }
   \label{fig1}
\end{figure}

\begin{figure*}[ht!]
  \includegraphics[width=\textwidth]{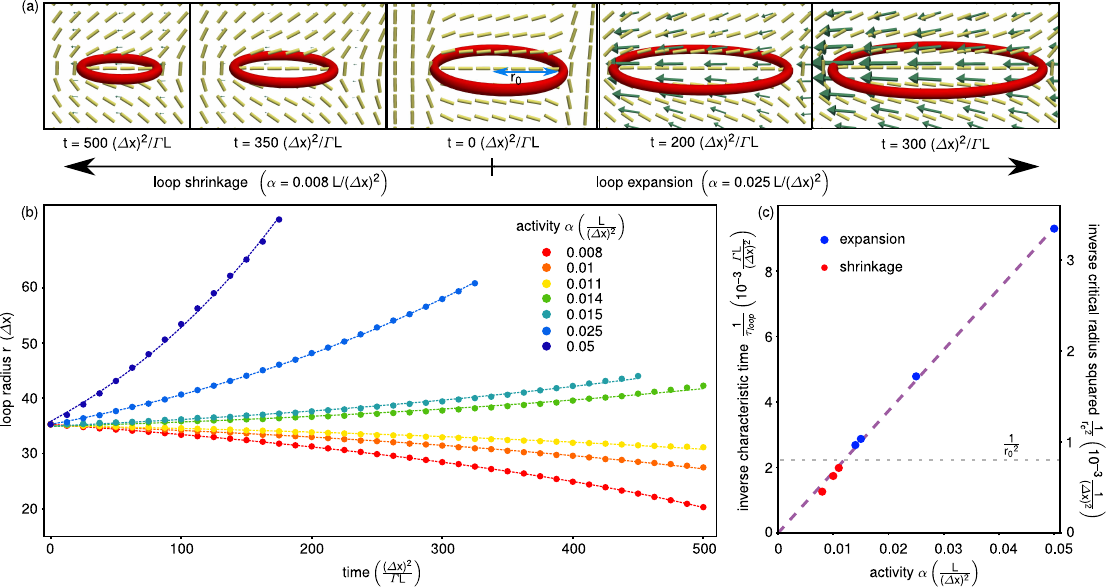}
  \caption{
  Shrinking and expanding dynamics of isolated active nematic defect loop. (a) Depending on activity $\alpha$, the defect loop either shrinks or expands in time; the panels show defect loop as isosurface of degree of order (red), director field (yellow rods), and velocity field (green arrows). The defect loop has $+1/2$ profile at the left side, and $-1/2$ profile on the right side, which generates an overall self propulsion velocity and a local active stress on the loop towards the left.
  (b) Single loop radius as function of time for different activities. Radius is determined as the loop left-to-right dimension in panel (a). Dashed line is a fit (Eq.~\ref{eq:r}) to the numerical data with fit parameters $\tau_\text{loop}$ and activity-independent parameter $r_\text{c}^2/\tau_\text{loop}$, obtaining $r_\text{c}^2/\tau_\text{loop}=2.78\,\Gamma L$.
  (c) Critical radius $r_\text{c}$ and initial radius $r_0=35.3\,\Delta x$ determine shrinkage ($r_0<r_\text{c}$) and expansion ($r_0>r_\text{c}$) regimes.
  A linear dependence of $1/\tau_\text{loop}$ (or equally $1/r_\text{c}^2$) on activity is obtained with the slope of $1/\tau_\text{loop}=0.186\,\Gamma\alpha$.
  }
  \label{fig:loop}
\end{figure*}

Coarsening of a defect network of three-dimensional active turbulence is demonstrated in Fig.~\ref{fig1}, following a quench from a high defect density regime.
The coarsening dynamics shows gradually decreasing defect density, and notably includes both shrinkage \emph{and} expansion of the length of  topological defect loops (Fig.~\ref{fig1}). 
We elucidate such shrinking and expansion dynamics by first considering the kinetics of isolated active defect loops that can be captured as the competition between the (elastic) line tension and the active propulsion.
For an inplane zero-topological charge loop of radius $r$ as shown in Fig.~\ref{fig:loop}a,
the defect line tension
can be estimated as 
$T=\frac{\pi K}{4}\ln\frac{r}{r_\text{min}}$, where $K$ is the single elastic constant proportional to tensorial elastic constant $L$ (see SI), 
and $r_\text{min}$ is the defect core size~\cite{KlemanM,MerteljA_PhysRevE69_2004}.
Such defect line also experiences an effective drag force due to local rotations of the director field as it moves through the material and can be estimated as $f_\text{drag}=c_\text{drag} v=\frac{\pi}{4}\gamma_1 v\ln\frac{r}{r_\text{min}}$, where $c_\text{drag}$ is the drag coefficient, $\gamma_1$ the rotational viscosity and $v$ its velocity with respect to the flow of the nematic fluid~\cite{KlemanM}.
The active self-propulsion flow velocity $v_0$ depends on the director field of different defect loop segments~\cite{GiomiL_PhilTransRSocA372_2014,BinyshJ_PhysRevLett124_2020} and varries from $v_0\approx 0$ for the $-1/2$ section to $v_0\approx\frac{|\alpha| r}{4\eta}$ for the $+1/2$ defect loop section.
If assuming a circular loop, all these contributions give a dynamical equation for the active nematic defect loop radius
\begin{equation}
\dot{r}=\frac{v_0}{2}-\frac{T}{c_\text{drag}\, r}=\frac{|\alpha| r}{8\eta}-\frac{K}{\gamma_1 r}.
\label{eq:rdot}
\end{equation}
Solving Eq.~\ref{eq:rdot} gives the time dependence of loop radius
\begin{equation}
    r(t)=r_\text{c}\left[1+\left(\frac{r_0^2}{r_\text{c}^2}-1\right)e^{t/\tau_\text{loop}}\right]^{1/2},
    \label{eq:r}
\end{equation}
where $\tau_\text{loop}=\frac{4\eta}{|\alpha|}$ is the characteristic time scale of isolated defect loops , $r_0$ is the initial loop radius at $t=0$, and 
$
    r_\text{c}=\sqrt{\frac{8\eta K}{\gamma_1|\alpha|}}
$
is the critical radius for which the active-propulsion exactly counterbalances the loop line tension (i.e. $\dot{r}=0$).  $r_\text{c}$ is explicitly dependent on nematic elasticity, activity and rotational viscosity, which provides a direct analytic insight into possible control of active defect loop kinetics; for $r>r_\text{c}$ the loops expand, whereas for $r<r_\text{c}$ the loops shrink.
Note that the existence of a critical radius has analogies with the spontaneous flow transitions in polar gels~\cite{VoituriezR_EurophysLett70_2005}, but with a notable difference that in polar gels the transition is symmetry breaking, whereas for active loops the flow is always generated and competes with the elasticity-induced shrinking.

The analytical model is compared to the full numerical simulation, observing excellent agreement (Fig.~\ref{fig:loop}). A loop with a fixed initial radius is let to dynamically evolve at different activities and depending on the activity, this leads to shrinking (Supplementary Movie 1) or expanding (Supplementary Movie 2) dynamics. A fit of Eq.~\ref{eq:r} to the simulation data gives $1/\tau_\text{loop}=0.186\,\Gamma\alpha$ and $r_\text{c}^2/\tau_\text{loop}=2.78\,\Gamma L$, which compares well to values of $1/\tau_\text{loop}=0.182\,\Gamma\alpha$ and $r_\text{c}^2/\tau_\text{loop}=2\,\Gamma L$ that are calculated directly from the viscoelastic parameters of the simulation.
More generally, now supported also by the numerical simulations, we show that the nematic elasticity, active propulsion and the viscous drag are the main mechanisms of the three-dimensional active defect kinetics.

\begin{figure*}[t]
  \includegraphics[width=\textwidth]{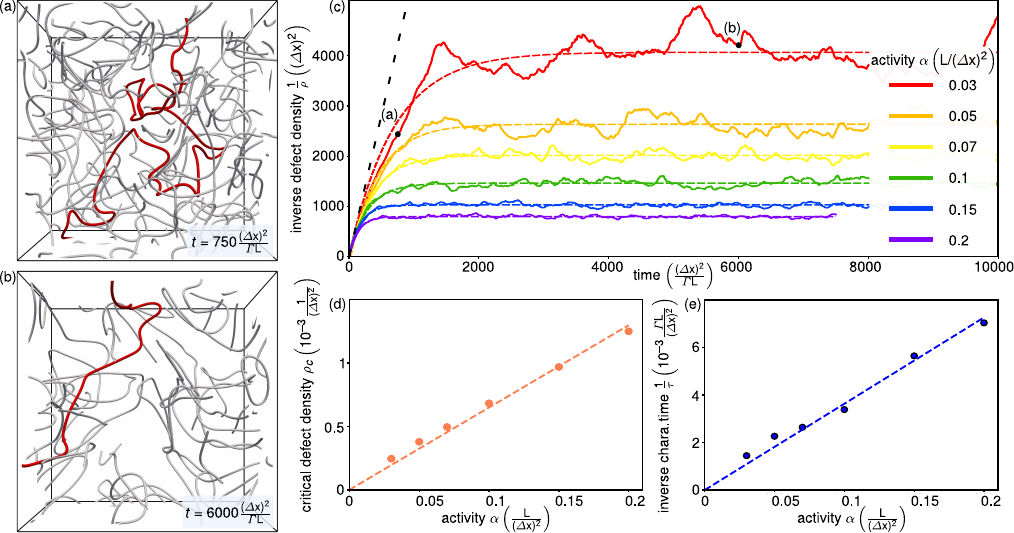}
  \caption{
  Coarsening dynamics of 3D active nematic turbulence.
  (a) Defect network at short time in the coarsening dynamics. A selected connected defect segment is colored in red.
  (b) Defect network at later time in the coarsening dynamics. Defect line segments are further apart and show smaller curvature compared to (a).
  (c) Inverse defect density over time for different activities. Dashed lines are fits of Eq.~\ref{eq:rho} to the simulation data. 
  Black dashed line represents the activity-independent initial density dynamics. The fluctuations of the defect density at later times is the effect of the finite simulation volume but with a well determined average.
  (d) Steady-state defect density as dependent on activity; linear fit (dashed line) has the slope of $0.0065/L$. 
  (e) Linear dependence of the inverse characteristic time on activity; the linear fit has coefficient of $0.036\,\Gamma$. Points in (d) and (e) are obtained from fits in (c). 
  }
  \label{fig:coarsening}
\end{figure*}

The mechanisms of defect line tension, drag force and self-propulsion that were used to describe the single active defects can be generalised to the overall coarsening dynamics of a full three-dimensional active defect network. Our model is based on a single time-dependent length scale $\xi$, which represents both typical radius of curvature and typical separation of defect lines~\cite{ChuangI_Science251_1991,KlemanM}. Notice that during active coarsening, both the average defect-defect separation and the curvature decrease over time (Fig.~\ref{fig:coarsening}a,b). Specifically, $\xi$ is calculated as $\rho=1/\xi^2$ where $\rho$ is the total defect length over unit volume. 
The coarsening dynamics of the defect networks is now described as the time evolution of the single length-scale $\xi$ based on a balance between the defect line tension $T/\xi$ and viscous drag $\Gamma\left(\dot{\xi}+b\,v_0\right)$, where the  phenomenological parameter $b$ describes the effective self-propulsion velocity of defect lines.
Generalized from a single defect loop (Eq.~\ref{eq:rdot}) to a defect network, here the line tension straightens and spaces out the defects in time, which gives the main dynamic equation of active coarsening
\begin{equation}
    \dot{\xi}=a\frac{K}{\gamma_1\xi}-b\frac{|\alpha|\xi}{4\eta},
    \label{eq:xidot}
\end{equation}
where in analogy to the passive coarsening~\cite{YurkeB_PhysRevE47_1993,WangW_JChemPhys108_1998}, we use dimensionless parameter $a$ to describe the relative strength of line tension compared to drag force in a defect network.
Equation~\ref{eq:xidot} can be rewritten in terms of the defect density
\begin{equation}
    \dot{\rho}=\frac{b|\alpha|}{2\eta}\rho-\frac{2aK}{\gamma_1}\rho^2,
    \label{eq:rhodot}
\end{equation}
and upon integration at constant activity, the coarsening equation for the active defect density is obtained
\begin{equation}
    \rho(t)=\rho_\text{c}\left[ 1+\left( \frac{\rho_\text{c}}{\rho_0}-1 \right)e^{-t/\tau} \right]^{-1},
    \label{eq:rho}
\end{equation}
where $\tau=\frac{2\eta}{b|\alpha|}$, $\rho_\text{c}=\frac{b|\alpha|\gamma_1}{4\eta aK}$, and $\rho_0$ is the initial density at $t=0$. 
Equation~\ref{eq:rho} shows how the defect density evolves from an initial value of $\rho_0$ towards a dynamic steady state $\rho_\text{c}$ with a well-defined time scale $\tau$.

Numerical modelling of coarsening for a full active nematodynamic approach
is shown in Fig.~\ref{fig:coarsening} and Supplementary Movie 3. The simulations are performed from an initial configuration of a random director field at each data point and  at $t\approx 10 \, \Delta x^2/(\Gamma L)$ a  dense defect network is formed, which coarsens over time (Fig.~\ref{fig:coarsening}c, Fig.~S1).
For high defect density at short times after quench, we observe that the coarsening dynamics is independent on activity, which can be explained by the elastic tension being much larger than the active self-propulsion in Eq.~\ref{eq:rhodot}.
At later times, the defect density approaches the steady-state density $\rho_\text{c}$, which we find is linearly proportional to activity (Fig.~\ref{fig:coarsening}d), in  full agreement with the analytical model. The rate of approach towards $\rho_\text{c}$ is governed by the time scale $\tau$, which is inversely proportional to activity (Fig.~\ref{fig:coarsening}e). 
Parameters $a$ and $b$ from the analytical model can now be determined by a linear fit in Figs.~\ref{fig:coarsening}(d,e), obtaining $a=2.8$ and $b=0.10$. 
Parameter $a$ is of roughly similar magnitude as in passive (i.e. zero activity) nematics~\cite{YurkeB_PhysRevE47_1993}, whereas a low value of $b$ indicates that the defects on average are repelled from each other with a much lower velocity than $v_0$, which is characteristic for isolated defect loops in Fig.~\ref{fig:loop}.
In the SI, we show that the coarsening dynamics is not significantly altered even for simulations with multiple nematic elastic constants (Fig. S4).

\begin{figure}[t]
  \includegraphics[width=\columnwidth]{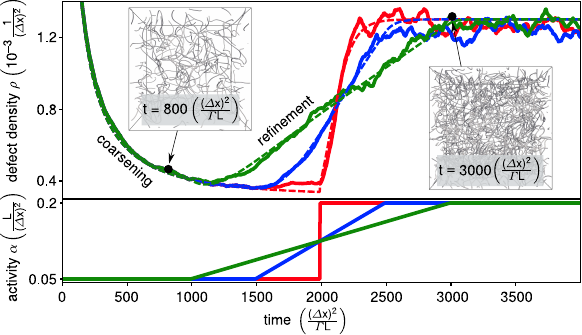}
  \caption{
  Active refinement induced by time-varying activity. Defect density (top graph) is shown for the three different time-varying activities (bottom graph). Note the matching colors. The dynamics is initiated at $t=0$ from a random director field, leading first to coarsening which upon an activity increase changes to refinement. Dashed lines are fits with Eq.~\ref{eq:rhodot}.
  }
    \label{fig4}
\end{figure}

Active refinement is --- oppositely to coarsening --- characterized by the proliferation of defects (for example induced by an increase of activity) and it
occurs in the regime where active propulsion prevails over the line tension.
Figure~\ref{fig4} and Supplementary Movie 4 show the defect density upon active refinement
from a full numerical simulation and in agreement with the
theoretical model (Eq.~\ref{eq:rhodot}).
Figure~\ref{fig4} also shows that the rate of the activity change affects the refinement dynamics, as it has to be compared to the characteristic time $\tau(t)$.
Fast changes in activity (red line in Fig.~\ref{fig4}) can  be described by constant activity dynamics (Eq.~\ref{eq:rho}), whereas for other regimes (blue and green lines in Fig.~\ref{fig4}) the full time-dependent activity $\alpha(t)$ has to be considered in Eq.~\ref{eq:rhodot}.
More generally, the results show that the introduced approach also well covers the \emph{time-dependent} changes in the active nematic material parameters,
such as activity, elasticity, and viscosity, indicating an exciting analytic insight into the kinetics of active states out of the dynamic equilibrium. Refinement could also be considered during a transition from an aligned initial condition to a defect network. We show such example in Fig. S5.

Experimentally, the demonstrated active coarsening (or refinement) could be induced by 
(meso)phase transitions into a nematic phase, 
triggered by a pressure or temperature quench~\cite{ChuangI_Science251_1991,AustinD_PhysRevD48_1993}, or possibly even by changing the activity as the nematic material parameters are known to be activity-dependent~\cite{ThampiSP_EPL112_2015,CatesME_AnnuRevCondensMatterPhys6_2015}.
The transition could be studied also in view of the Kibble-Zurek mechanism, which is known to describe structure formation in liquid crystals~\cite{BradacZ_JChemPhys135_2011,FowlerN_ChemPhysChem18_2017}, and could interestingly be additionally coupled to curved interfaces~\cite{StoopN_SoftMatter14_2018} and topology of the confining space~\cite{NikkhouM_NaturePhys11_2015}.

The demonstrated coarsening and refinement of active defects also shows  interesting implications beyond soft and active matter.  Distinctly, at zero activity, the coarsening dynamics of nematic defect strings is known to share strong mathematical similarities with cosmic strings and condensed matter systems~\cite{KibbleTWB_JPhysAMathGen9_1976,ZurekW_PhysRep276_1996}, whereas here we show that activity can contribute new coarsening terms.
Namely, in an active nematic system, the time derivative of the characteristic length scale $\xi$ (Eq.~\ref{eq:xidot}) equals an elastic term proportional to $1/\xi$ and an active term proportional to $\xi$. This coarsening dynamics shows an  interesting analogy with the velocity-dependent one-scale cosmic string model~\cite{MartinsCJAP_PhysRevD65_2002,MartinsCJAP_PhysRevD94_2016}, where
(i) the friction term due to particle scattering proportional to $1/\xi$ is known to give the Kibble coarsening scaling $\xi\sim t^{0.5}$,
and (ii) the term proportional to $H\xi$, where $H$ is the Hubble parameter, accounts for the expansion of the universe. Both terms in the cosmic string model are positive and promote the coarsening dynamics, whereas for active nematics the $\xi$ term is negative ($\dot{\xi}\sim-\xi$) and as we demonstrate can slow down the coarsening and leads to a dynamic steady-state with a finite defect density. Such a term would correspond to coarsening of a string network in a shrinking universe. More generally, this   analogy provides novel unprecedented parallels between active matter and cosmology.

\begin{acknowledgments}
The authors acknowledge funding from Slovenian Research Agency (ARRS) under contracts P1-0099, N1-0124, J1-1697, J1-2462, N1-0195, and from European Research Council  grant LOGOS. 
\end{acknowledgments}


\widetext
\clearpage

\setcounter{figure}{0}
\renewcommand\vec{\boldsymbol}
\renewcommand{\thefigure}{S\arabic{figure}}

\section{Supplemental Material}
\section{Analytical model of defect loop dynamics}

An effective model of the shrinking and expansion dynamics of active nematic loops is formulated, considering the effects of elastic line tension, self-propulsion of active defect line segments, and viscous drag on moving defect lines. The deformation dynamics of active loops depends on the geometry of the director field profile~\cite{BinyshJ_PhysRevLett124_2020}. The model is able to capture the dynamics of loops that remain within a plane (e.g. defect loops in Fig. 2) and do not buckle out of the plane in time.

The shrinking of defect loops is driven by the defect line tension that describes the elastic energy of the director field per length of a defect line segment and can be written as~\cite{KlemanM}
\begin{equation}
    T=\frac{\pi K}{4}\ln\frac{r_\text{max}}{r_\text{min}},
\end{equation}
where $K$ is the single elastic constant and $r_\text{min}$ is the defect core size. $r_\text{max}$ represents the radial size of the region effectively associated with the defect line. It is typically related to defect-defect separations or the radius of curvature. For circular loops with radius $r$, we set $r_\text{max}=r$.

The defect lines are singularities in the orientational field of the nematic fluid and their movement relative to the fluid flow corresponds to reorientation of the nematic director field. Director rotation dissipates the energy with the dissipation rate $\Sigma=\frac{\pi}{4}\gamma_1 v^2\ln\frac{r_\text{max}}{r_\text{min}}=c_\text{drag} v^2$~\cite{KlemanM}, where $\gamma_1$ is the rotational viscosity and $v$ is the velocity magnitude of the half-integer defect line in its normal plane and relative to the fluid flow. We defined the drag coefficient $c_\text{drag}=\frac{\pi}{4}\gamma_1 \ln\frac{r_\text{max}}{r_\text{min}}$ and take for circular loops similarly as for the elastic tension $r_\text{max}=r$.

The dynamics of defect lines can be captured by considering a balance between energy relaxation and energy dissipation, which can be expressed as a balance of effective forces acting on defect line segments. For passive circular defect lines, the line density of the elastic tension force equals $T/r$ and is counterbalanced by the drag force density $c_\text{drag}\dot{r}$, which leads to a dynamic equation for shrinking loops. In our active loop model, we consider the same principles of elastic tension force and drag force, but in a presence of material flow, generated by activity.

The active propulsion is estimated by using self-propulsion velocities $v_0$ for different defect loop segments~\cite{GiomiL_PhilTransRSocA372_2014,BinyshJ_PhysRevLett124_2020}; for the $+1/2$ defect loop section, we take $v_0\approx\frac{|\alpha| r}{4\eta}$, where $\alpha$ is the activity and $\eta$ the effective isotropic viscosity, whereas for the $-1/2$ section we assume no self-propulsion.
The intermediate parts of the loop with twist director profiles also experience self-propulsion~\cite{BinyshJ_PhysRevLett124_2020}. For simplicity, our model assumes a circular loop. Self-propulsion velocity of each loop segment would become important in models that would consider also the evolution of the shape of the loop in time.
The considered self-propulsion velocities are relevant for curved defect lines in three dimensions. For perfectly straight defect lines, possible friction effects and the system size can also become important~\cite{GiomiL_PhilTransRSocA372_2014}.

To derive the dynamics of the loop size, we consider the force balance on the far left and the far right parts of the loop with positions $x_\text{l}$ and $x_\text{r}$, respectively (Fig. S1). In the model of a circular loop, the radius is determined as $r=(x_\text{r}-x_\text{l})/2$. The left part of the loop is advected by the flow field $v_0$. The displacement of the left part is given by the force balance 
\begin{equation}
\frac{T}{r}=c_\text{drag}\left( \dot{x}_\text{l}+v_0 \right),
\end{equation}
where $v_0=\frac{|\alpha| r}{4\eta}$. The right part of the loop generates no self-advection and obeys the equation
\begin{equation}
\frac{T}{r}=-c_\text{drag} \dot{x}_\text{r}.
\end{equation}
Finally, we obtain a dynamical equation for the active nematic defect loop radius
\begin{equation}
\dot{r}=\frac{ \dot{x}_\text{r}- \dot{x}_\text{l}}{2}=\frac{v_0}{2}-\frac{T}{c_\text{drag}\, r}=\frac{|\alpha| r}{8\eta}-\frac{K}{\gamma_1 r}.
\label{eq:rdot}
\end{equation}
Solving Eq.~\ref{eq:rdot} gives the time dependence of loop radius
\begin{equation}
    r(t)=r_\text{c}\left[1+\left(\frac{r_0^2}{r_\text{c}^2}-1\right)e^{t/\tau_\text{loop}}\right]^{1/2},
    \label{eq:r}
\end{equation}
where $\tau_\text{loop}=\frac{4\eta}{|\alpha|}$ is the characteristic time scale of isolated defect loops, $r_0$ is the initial loop radius at $t=0$, and 
$
    r_\text{c}=\sqrt{\frac{8\eta K}{\gamma_1|\alpha|}}
$
is the critical radius for which the active-propulsion exactly counterbalances the loop line tension (i.e. $\dot{r}=0$).  For $r>r_\text{c}$ the self-propulsion prevails and the loops expand, whereas for $r<r_\text{c}$ the line-tension prevails and the loops shrink. The model allows also for geometries of the director field around the defect loop that are different to Fig. 2, as long buckling modes are suppressed. For example, the $+1/2$ segment of the defect loop could point towards the $-1/2$ segment, in which case the activity would speed up the shrinking. Likewise, we could predict the shrinking and growing dynamics for non-buckling defect loops with a non-zero topological charge.

\begin{figure*}[h!]
  \includegraphics[width=6cm]{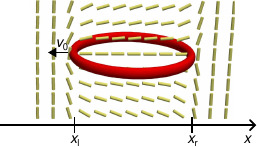}
  \caption{
  Active nematic defect loop. The dynamical model considers the force balance on the far left and the far right part of the loop with positions $x_\text{l}$ and $x_\text{r}$, respectively. 
  }
  \label{fig0}
\end{figure*}

\section{Details on numerical modeling}

We perform extensive mesoscale simulations of active nematics based on the Beris-Edwards approach to nematodynamics extended with the active stress tensor~\cite{DoostmohammadiA_NatCommun9_2018,CoparS_PhysRevX9_2019,CarenzaLN_ProcNatlAcadSci116_2019,ZhangR_NatCommun7_2016,HatwalneY_PhysRevLett92_2004}.
The approach is based on 
a set of coupled dynamic equations for the orientational order (given by the $\mathrm{Q}$-tensor) and the fluid velocity $\vec{v}$. 
The $\mathrm{Q}$-tensor evolves as
\begin{equation}\label{eq:q_tensor}
    \left(\partial_t + v_k \partial_k \right) Q_{ij} - S_{ij} = \Gamma H_{ij},
\end{equation}
where $\vec{v}$ is the fluid velocity and $\Gamma$ is the rotational viscosity coefficient. The generalized advection term $S_{ij}$ includes the effects of velocity gradients on the nematic order 
\begin{equation}
S_{ij} = \left(\chi D_{ik} -\Omega_{ik} \right)\left(Q_{kj} + \frac{1}{3}\delta_{kj} \right) + \left(Q_{ik} + \frac{1}{3}\delta_{ik} \right)\left(\chi D_{kj} +\Omega_{kj} \right) - 2\chi \left(Q_{ij} + \frac{1}{3}\delta_{ij}\right) Q_{kl} W_{lk},
\end{equation} 
where $\chi$ is the flow alignment parameter, $D_{ij} = \frac{1}{2}(\partial_i v_j+\partial_j v_i)$ and $\Omega_{ij} = \frac{1}{2}(\partial_i v_j - \partial_j v_i)$ are symmetric and antisymetric part of the velocity gradient tensor $W_{ij}=\partial_i v_j$, respectively, and $H_{ij} = - \delta F/Q_{ij} + \frac{1}{3}\delta_{ij} \mathrm{Tr} (\delta F / \delta Q_{ij})$ is the molecular field that drives system towards the equilibrium of the free energy $F$. The free energy is written in the Landau-de Gennes single elastic form as
\begin{equation}
F = \int \left( \frac{A}{2}Q_{ij}Q_{ji} + \frac{B}{3}Q_{ij}Q_{jk}Q_{ki} + \frac{C}{4}\left( Q_{ij} Q_{ji}\right)^2 + \frac{L}{2} \left(\partial_k Q_{ij} \right)^2 \right)  \mathrm{d}V,
\end{equation}
where $A$, $B$ and $C$ are material parameters and $L$ is the elastic constant. 
The flow field obeys the continuity equation $\partial_t \rho + \partial_i  \left( \rho v_i \right) = 0$ and the Navier-Stokes equation $\rho \left(\partial_t + v_j \partial_j \right) v_i = \partial_j \Pi_{ij}$, where $\rho$ is the fluid density and $\Pi_{ij}$ is the stress tensor, consisting a passive and an active term $\Pi_{ij} = \Pi_{ij}^\text{passive} +\Pi_{ij}^\text{active}$,
\begin{align}
       \Pi_{ij}^\text{passive} = &-p \delta_{ij} + 2\chi \left( Q_{ij} +\frac{1}{3} \delta_{ij} \right) Q_{kl} H_{kl} - \chi H_{ik} \left( Q_{kj} +\frac{1}{3} \delta_{kj} \right) - \chi \left( Q_{ik} +\frac{1}{3} \delta_{ik} \right)H_{kj}+\nonumber\\
       &+ Q_{ik} H_{kj} - H_{ik} Q_{kj} + 2 \eta D_{ij} - \partial_i Q_{kl}\frac{\delta F}{\delta \partial_j Q_{kl}},
\end{align}
\begin{equation}
    \Pi_{ij}^\text{active} = -\alpha Q_{ij} ,
\end{equation}
where $p$ is the pressure, $\eta$ is the isotropic viscosity and $\alpha$ is the activity, which is positive in extensile materials and negative in contractile materials. The coupled equations for the nematic order and the fluid velocity are solved numerically using the hybrid lattice-Boltzmann approach~\cite{CoparS_PhysRevX9_2019,CarenzaLN_ProcNatlAcadSci116_2019,ZhangR_NatCommun7_2016}. This approach consists of finite difference method for solving the Q-tensor evolution (Eq. \ref{eq:q_tensor}),  and the D3Q19 lattice Boltzmann method for the  Navier-Stokes equation and the continuity equation.\par

The simulations of loop shrinkage and expansion were performed on a $300 \times 300 \times 300$ mesh size, while the coarsening dynamics were obtained on a $400 \times 400 \times 400$ mesh. Periodic boundary conditions were used in all the directions of the simulation box. Mesh resolution is defined as $\Delta x = 1.5 \xi_n$, where  $\xi_n$ is nematic correlation length $\xi_n = \sqrt{L/(A+BS_{eq}+\frac{9}{2} C S_{eq}^2)}$ and $S_{eq}$ is the equilibrium value of the scalar order parameter.
The results of the simulations are expressed in the units of the mesh resolution $\Delta x$, rotational viscosity parameter $\Gamma$ and elastic constant $L$. 
The time step in case of loop dynamics is set to $\Delta t = 0.0025\,(\Delta x)^2/(\Gamma L) $. In the coarsening and refinement dynamics the time step equals $\Delta t =0.025\,(\Delta x)^2/(\Gamma L) $.
The following values of the model parameters are used: $A = -0.43 L/(\Delta x)^2$, $B = -5.3 L/(\Delta x)^2$, $C = 4.325 L/(\Delta x)^2$, $S_{eq} = 0.533$, $\eta=1.38/\Gamma$, and $\chi=1$. Elasticity $L$ and viscosity $\Gamma$ parameters in the Q-tensor formulation of nematodynamics can be expressed with the parameters $K$ and $\gamma_1$ from the director formulation as used in Eq.~1 using the relations: $\gamma_1=9S^2/(2\Gamma)$ and $K=9LS^2/2$, where $S$ is the scalar order parameter.
Elastic constant $K$ is dependant on the scalar degree of order $S$. Our analytical approaches use the directorial description of nematodynamics and use the $K$ elastic constant computed at $S=S_\text{eq}$, while the numerical simulation use a tensorial approach with the $L$ elastic constant.\par

In simulations we consider a single elastic constant $L_1$, except in Fig.~\ref{fig:rho_L21} where we use an elastic free energy density with two nonzero elastic constants $L_1$ and $L_2$:
\begin{equation}
f_\text{elastic}=\frac{L_1}{2} \left(\partial_k Q_{ij} \right)^2 + \frac{L_2}{2} \left(\partial_i Q_{jk} \right)\left(\partial_j Q_{ik} \right).
\end{equation}
Using two elastic constants results in different elastic constants for twist distortions ($K_2$) compared to splay and bend distortions ($K_1=K_3$). Additionally, using $L_1$ and $L_2$ also changes the nematic correlation length, which is importantly linked to the resolution of our numerical mesh. To address this issue, we have computed the average elastic constant of the director distortions $\bar{K}=(K_1+K_2+K_3)/3$ and compute the effective elastic constant in the tensorial approach $L'=2\bar{K}/(9S_\text{eq}^2)= L_0( \frac{3L_1 + L_2}{3L_1})$, where $L_0$ represents the constant value of previous normalization term $L$.  We also considered the connection $L_2 = \beta L_1$, where $\beta = 0$ in case of single elastic constant, which gives the relation $L' = L_0 ( 1+ \beta/3)$. The results of the simulations using different values of $L_1$ and $L_2$ are in Fig. S5 expressed in units of $L_0$ and $\Delta x_0=1.5\sqrt{L'/(A+BS_{eq}+\frac{9}{2} C S_{eq}^2)}$, which remain constant for all simulations. Such approach allows us the compare simulations performed at different values of $L_1$ and $L_2$.

\newpage

\section{Supplementary Figures}

\begin{figure*}[h!]
  \includegraphics[width=13cm]{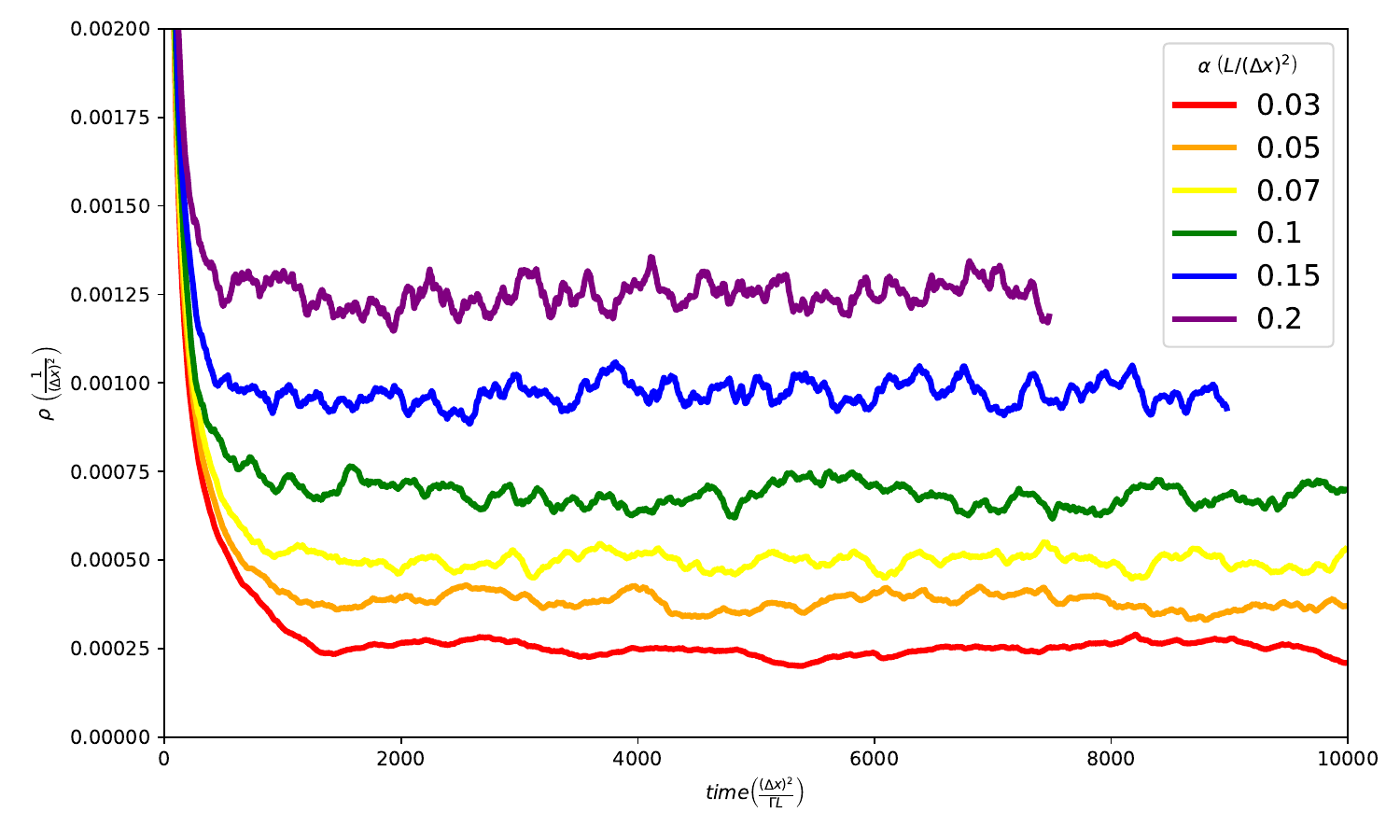}
  \caption{Coarsening dynamics of three-dimensional active nematic defect networks. Defect density is shown over time for different activities. The numerical simulation is started from random director field initial condition. The coarsening leads to activity-dependent dynamic steady states, described by the steady state defect density $\rho_\text{c}$.
  The defect density is defined as the length of defect lines over a unit volume and is computed from the defect volume fraction that describes defect regions where scalar order parameter is $S < 0.4$.
  }
  \label{fig:rho}
\end{figure*}

\begin{figure*}[h!]
  \includegraphics[width=13cm]{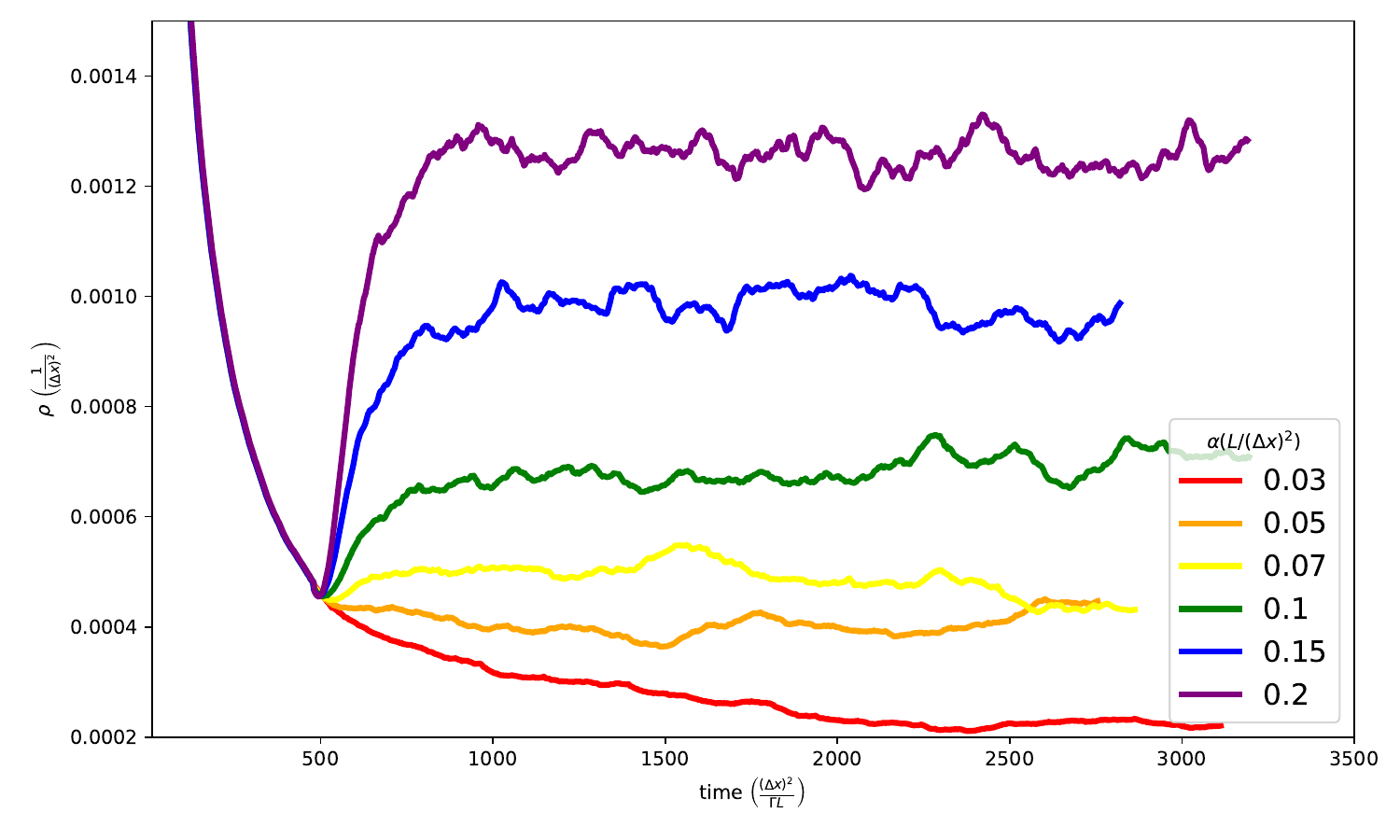}
  \caption{Active coarsening and refinement for three-dimensional active nematics. The coarsening dynamics is started from a random director field initial condition at zero activity. At $t=500\,(\Delta x)^2/(\Gamma L)$, the activity is increased to a finite value. If the activity is increased to a small value, the coarsening dynamics is slowed down. At higher activity increase, the coarsening dynamics turns to refinement and the defect density is increased. 
  }
  \label{fig:rho_alpha_increase}
\end{figure*}

\begin{figure*}[h!]
  \includegraphics[width=12cm]{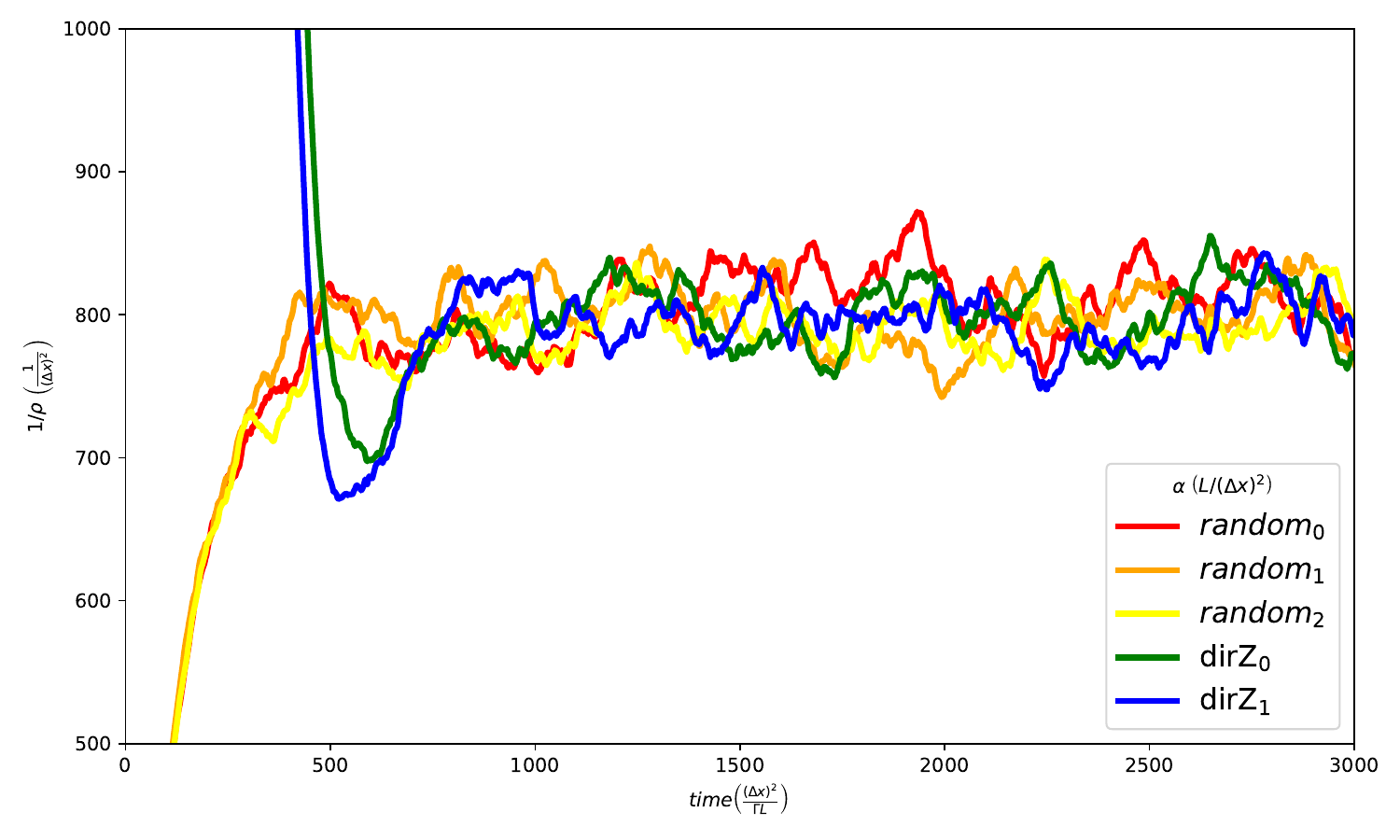}
  \caption{
  Coarsening dynamics of three-dimensional active nematic defect networks at activity $\alpha = 0.2 L/(\Delta x)^2$. The numerical simulations started from different initial conditions of director field. The simulations indicated with $random_{*}$ had started from different random director field, while the other indicated with $dirZ_{*}$ had started from director field aligned in z-direction with added noise. The coarsening dynamics depends on the initial condition of the director field, however the steady state $\rho_\text{c}$ stays the same. The comparison of the dynamics is shown also in Supplementary Movie 5.}
  \label{fig:rho_IC}
\end{figure*}

\begin{figure*}[h!]
  \includegraphics[width=12cm]{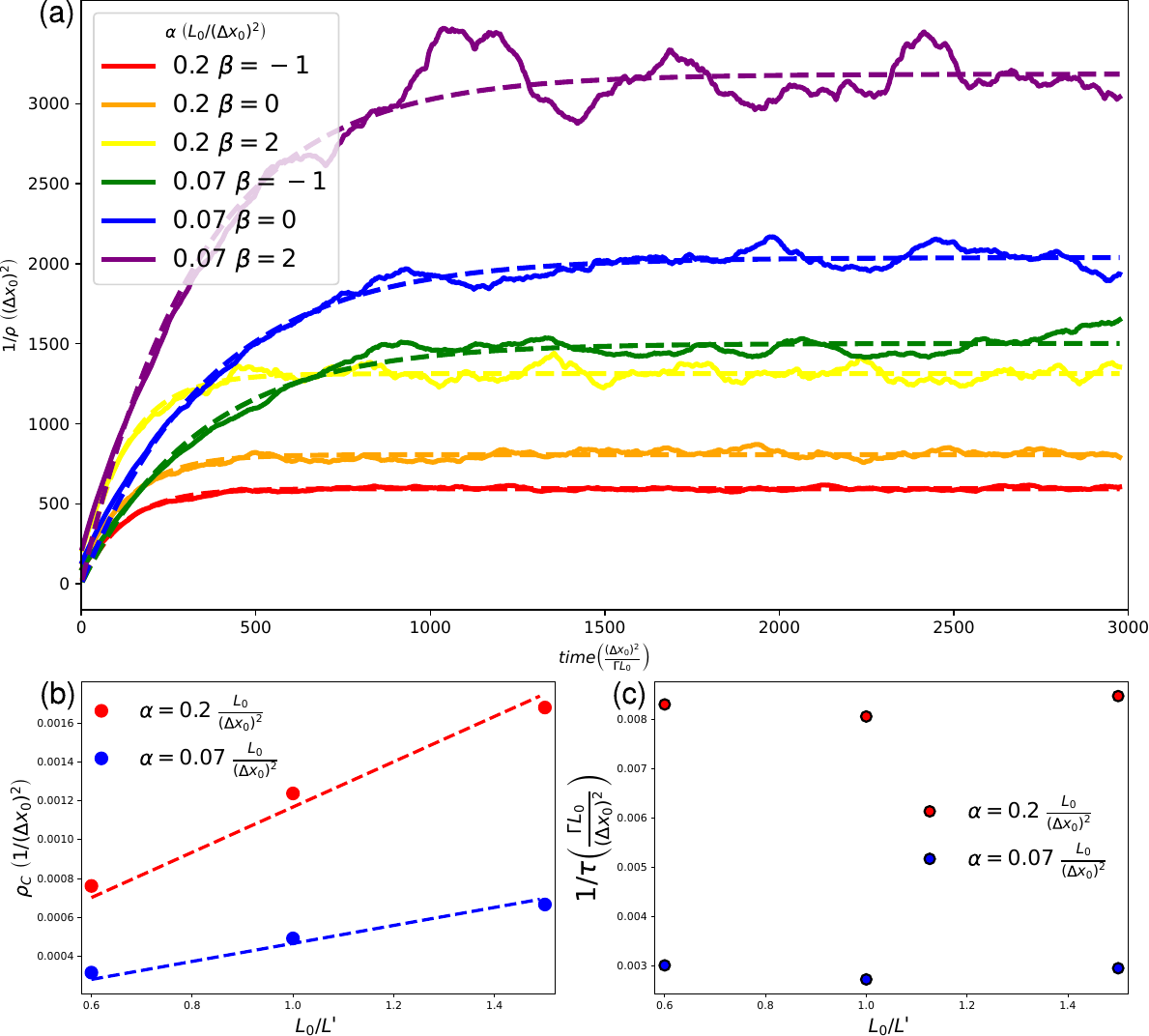}
  \caption{
  Coarsening dynamics of three-dimensional active nematic defect networks for multiple elastic constants and two different activities. Extra elastic term affects the overall dynamics of the system, including the steady-state defect density $\rho_c$; however, we observe that the behaviour of $\rho_c$ at different elastic constants  $L_1$ and $L_2$ can be described by considering only one effective elastic constant $L'$. As expected, the characteristic time $\tau$ is mostly independent on the elastic constants.
  }
 
  \label{fig:rho_L21}
\end{figure*}

\end{document}